\pdfoutput=1
\documentclass[11pt]{article}
\usepackage[affil-it]{authblk} 
\usepackage[authoryear]{natbib}
\usepackage{gensymb}
\usepackage{amsmath}
\usepackage{amssymb}
\usepackage{graphicx}
\usepackage{float}
\usepackage{booktabs}
\usepackage[a4paper, top=1in,bottom=1in,left=1in,right=1in,marginparwidth=0cm]{geometry}
\usepackage[title]{appendix}
\usepackage{pdflscape}
\usepackage{longtable}
\usepackage{adjustbox}
\usepackage{array}
\usepackage{caption}
\usepackage{chngcntr}
\usepackage{mathtools}
\usepackage{lineno}
\usepackage{setspace}
\usepackage{makecell}
\usepackage{hyperref}
\hypersetup{colorlinks=true,allcolors=blue}
\graphicspath{{figures/}}

\newcolumntype{R}[2]{%
    >{\adjustbox{angle=#1,lap=\width-(#2)}\bgroup}%
    l%
    <{\egroup}%
}

\renewcommand{\eqref}[1]{Eq.\,\ref{#1}}

\title{Stratified dispersal explains mountain pine beetle's range expansion in Alberta}

\author[1,*]{Evan C. Johnson}
\author[2]{Micah Brush}
\author[2,3]{Mark A. Lewis}
\affil[1]{Mathematical and Statistical Sciences; University of Alberta; Edmonton, Alberta, Canada}
\affil[2]{Department of Biology; University of Victoria; Victoria, British Columbia, Canada}
\affil[3]{Department of Mathematics and Statistics; University of Victoria; Victoria, British Columbia, Canada}
\affil[*]{Corresponding author: Evan Johnson, ecjohns1@ualberta.ca}

\date{}

\begin{document}

\maketitle

\newpage

\tableofcontents

\newpage

\section*{Abstract}


\begin{enumerate}

\item The mountain pine beetle (MPB), a destructive pest native to Western North America, has recently extended its range into Alberta, Canada. 

\item Predicting the dispersal of MPB is challenging due to their small size and complex dispersal behavior. Because of these challenges, estimates of MPB's typical dispersal distances have varied widely, ranging from 10 meters to 18 kilometers.

\item Here, we use high-quality data from helicopter and field-crew surveys to parameterize a large number of dispersal kernels. 

\item We find that fat-tailed kernels --- those which allow for a small number of long-distance dispersal events --- consistently provide the best fit to the data. Specifically, the radially-symmetric Student's $t$-distribution with parameters $\nu = 0.012$ and $\rho = 1.45$ stands out as parsimonious and user-friendly; this model predicts a median dispersal distance of 60 meters, but with the $95^\text{th}$ percentile of dispersers travelling nearly 5 kilometers. 

\item The best-fitting mathematical models have biological interpretations. The Student's t-distribution, derivable as a mixture of diffusive processes with varying settling times, is consistent with observations that most beetles fly short distances while few travel far; early-emerging beetles fly farther; and larger beetles from larger trees exhibit greater variance in flight distance.

\item Finally, we explain why other studies have found such a wide variation in the length scale in MPB dispersal, and we demonstrate that long-distance dispersal events are critical for modelling MPB range expansion.
\end{enumerate}

\textbf{Keywords:} mountain pine beetle, long distance dispersal, stratified dispersal, range expansion, bark beetle, fat-tailed distribution, dispersal kernel

\pagebreak


\section{Introduction}

Dispersal is the key factor driving the spread of forest pests. The assumption of simple random motion by individuals leads to a Gaussian distribution for dispersal distances. Although there is tremendous variability in dispersal data, kernels tend to be leptokurtic, having more long and short-distance values than found in a Gaussian distribution with comparable variance~\citep{kot_dispersal_1996}. The spread rate of a biological invasion is critically sensitive to the shape of the dispersal kernel~\citep{lewis_mathematics_2016}.  In particular, long-distance dispersal can increase the spread rate by orders of magnitude.  In the extreme case of fat-tailed dispersal kernels (those not exponentially bounded like the thin-tailed Gaussian distribution), invasions may continually increase in speed and never achieve a constant spread rate. 

The mountain pine beetle (MPB), \textit{Dendroctonus ponderosae} Hopkins (\textit{Coleoptera: Curculionidae}), exemplifies the critical role of dispersal in forest pest invasions. The most recent outbreak of MPB is the largest bark beetle outbreak ever recorded, killing pine trees across western North America and affecting 18 million hectares in British Columbia alone \citep{taylor2006forest, walton2012provincial}. Driven by fire suppression and climate change \citep{carroll2006impacts, creeden2014climate}, MPB expanded its historic range in a northeasterly direction, breaching the Rocky Mountains to reach Alberta \citep{nealis2014risk, carroll2003bionomics, bleiker_risk_2019}. Although MPB has currently stalled in eastern Alberta (putatively due to some combination of control efforts, host-tree depletion, and unusually cold winters; Janice Cooke qtd. in \citealp{cbc2023mpb}), future outbreaks may invade pine forests stretching to the east coast of North America.

MPB's recent range expansion into Alberta offers a unique opportunity to study dispersal dynamics due to two key factors: data availability and the absence of endemic populations. MPB uses aggregation pheromones to coordinate mass attacks on pine trees, resulting in rapid drying out of the trees and rust-red foliage. This clear diagnostic feature, along with MPB's economic importance \citep{corbett2016economic}, has led to extensive data collection via aerial surveys. Alberta also lacks endemic populations of MPB, thus eliminating a significant confounding factor. Endemic populations are practically unobservable due to low densities (around 40 individuals per hectare; \citealp{carroll2006mountain}) and because they do not produce the diagnostic red-topped trees; instead of mass-attacking healthy trees, endemic beetles persist in dead or severely weakened trees with a suite of other bark beetles. Therefore, in regions with endemic populations, it is often unclear whether red-topped trees result from endemic beetles becoming able to attack healthy trees (due to climatic conditions or recent disturbances; \citealp{bleiker2014characterisation}), or from epidemic beetles dispersing from elsewhere. 

Describing the flight-based dispersal of MPB is key to describing their spread, yet this task is confounded by biological complexity, as well as our inability to track individual beetles. The dispersal of MPB is influenced by various environmental factors, including wind, temperature, and host-tree availability \citep{chen_climatic_2017, mccambridge_temperature_1971, powell_phenology_2014}. At shorter scales, their ability to aggregate and overcome host defenses is facilitated by chemical signaling \citep{raffa_mixed_2001}. Once a tree has been ``filled up'' with MPB, additional beetles are deterred by anti-aggregation pheromones and auditory signals (i.e. stridulations). More importantly, MPB display distinct dispersal behaviors: they can fly short distances through the forest to find suitable hosts, or they can disperse long distances \textit{en masse} above the canopy, carried by convective wind currents. Both modes of dispersal are important at the landscape scale, but long-distance dispersal events are relatively rare and thus difficult to predict. Existing models of MPB often utilize dispersal kernels (e.g., Gaussian or Laplace distributions) that implicitly ignore long-distance dispersal events, even though these events determine the speed of range expansion \citep{liu2019accelerating}. 

The existing literature finds significant variation in the typical length scale of MPB dispersal (e.g., the median dispersal distance), from 10 m, to 1 km, to 10s of km (see Table \ref{tab.data}). Statistical models of dispersal --- which use covariates like ``total infestations within 3 km'' rather than dispersal kernels --- similarly find that typical dispersal distances vary over several orders of magnitude \citep[e.g.,][]{preisler_climate_2012,sambaraju_climate_2012,powell_phenology_2014}. This degree of variation cannot be attributed to true spatial or temporal heterogeneity in beetle dispersal. Instead, the wide range of estimates is an artifact of questionable modeling choices (see Discussion).

\renewcommand{\arraystretch}{2}
\begin{table}[H]
\caption{
A summary of existing literature on beetle dispersal. We present the study, the characteristic scale of dispersal, and the methodology used. We additionally include the most direct estimates (those from mark-recapture or nearest neighbor studies) for other related bark beetles. Additional information on each of these sources and how we obtained the values in this table are available in Appendix \ref{app:length}.}
\label{tab.data}
    \centering
    \begingroup\fontsize{8pt}{9pt}\selectfont
    \noindent\makebox[\textwidth]{%
    \begin{tabular}{l l l l}
        \toprule \toprule
        Study & Scale (m) & Method \\ \midrule \midrule
        \citet{aukema_movement_2008} & 18 000 & Statistical model & \\
        \citet{koch_signature_2021} & 17 000 & Dispersal kernel &\\
        \citet{preisler_climate_2012} & 10 000 & Statistical model &\\
        \citet{sambaraju_climate_2012} & 6000 & Statistical model &\\
        \citet{howe_climate-induced_2021} & 5000 & Statistical model &\\
        \citet{carroll_assessing_2017} & 2000 & Nearest neighbor &\\
        \citet{simard_what_2012} & 2000 & Statistical model &\\
        \citet{robertson_spatialtemporal_2009} & 1000 & Modified nearest neighbor &\\
        \citet{strohm_pattern_2013} & 364 & Other model &\\
        \citet{powell_phenology_2014} & 5 -- 90 & Other model &\\
        \citet{robertson_mountain_2007} & 30 -- 50 & Nearest neighbor &\\
        \citet{safranyik_dispersal_1992} & 30 & Mark-recapture &\\
        \citet{heavilin_novel_2008} & 10 -- 15 & Dispersal kernel &\\ 
        \citet{goodsman_aggregation_2016} & 10 & Dispersal kernel &\\
        \midrule
        \multicolumn{3}{l}{Data estimates from related bark beetles} & Bark beetle species\\
        \midrule
        \citet{withrow_spatial_2013} & 1000 -- 2500 & Nearest neighbor & Douglas-fir beetle \\
        \citet{turchin_quantifying_1993} & 690 & Mark-recapture & Southern pine beetle \\
        \citet{werner_dispersal_1997} & 90 -- 300 & Mark-recapture & Spruce beetle \\
        \citet{zumr_dispersal_1992} & 200 & Mark-recapture & European spruce beetle \\
        \citet{dodds_sampling_2002} & 200 & Mark-recapture & Douglas-fir beetle \\
        \citet{kautz_quantifying_2011} & 100 & Nearest neighbor & European spruce beetle \\ 
        \citet{zolubas_recapture_1995} & 10 & Mark-recapture & European spruce beetle \\
        \bottomrule
    \end{tabular}
}
\endgroup
\end{table}

There have also been attempts to characterize beetle dispersal without an underlying model. A mark-recapture experiment with MPB found typical dispersal distances of around 30 m \citep{safranyik_dispersal_1992}, and other mark-recapture experiments with related beetles find similarly short distances (see Table \ref{tab.data}). A notable shortcoming of these experiments is that they utilize pheromone traps, which may bias dispersal estimates downwards. Another approach that uses the data directly involves finding new infestations and assigning them to the closest infestation in the previous year. For brevity, we call these \textit{nearest neighbor studies} in Table \ref{tab.data}. Like the mark-recapture studies, nearest-neighbor studies suffer from downward bias, and like the dispersal kernel studies, the nearest-neighbor studies find a large range of typical dispersal distances.

Previous studies do not agree on the scale of MPB dispersal, and it can be difficult to judge individual studies without expertise in both MPB biology and modeling; therefore, there is a need for a reliable, authoritative, and easy-to-use dispersal model. In addition to predicting MPB infestations in the short term, describing MPB dispersal is a key stepping stone to answering longstanding applied questions. These include 1) how spread is affected by host resistance, which is thought to vary both within and across pine species \citep{cudmore_climate_2010, six_are_2018, srivastava_dynamic_2023}; 2) the relationship between infestation density and dispersal \citep{safranyik_biology_2006, jones_factors_2019}; 3) the risk of MPB invasion into the low-volume Jack pine stands of Eastern Alberta and Saskatchewan; and 4) the risk of MPB invasion into the high-volume stands of Northwest Ontario \citep{bleiker_risk_2019}.

In this paper, we parameterize dispersal kernels using high-quality infestation data from the Government of Alberta. The intensive survey effort undertaken provides data on the number of infested trees throughout Alberta with a positional accuracy of $\pm$ 30 m. Importantly, because beetles were not previously present in Alberta, new outbreak locations are not thought to be from erupting endemic beetle populations, but instead from dispersal events. We consider a large class of possible kernels, including thin and fat-tailed distributions. We anticipate that given the importance of long-distance dispersal events in biological invasions, fat-tailed distributions should provide the best fit to the infestation data. 

\section{Materials and methods}

\subsection{Background}

Beetles emerge from their natal trees and attack new host trees each summer. If a tree is killed, its foliage will turn rust-red the proceeding summer (or early autumn), approximately one year after the initial attack. Knowing this diagnostic feature of year-old infestations, the Alberta Department of Agriculture and Forestry uses helicopters to find all red-topped trees across vast swaths of Alberta’s forests. These so-called Heli-GPS surveys are an integral part of Alberta’s management strategy for mountain pine beetle. After the heli-GPS surveys are conducted each year (typically around September), field crews are sent to the locations of red-topped trees. The crews then perform concentric ground surveys to find nearby green-attack trees: trees that were recently infested and containing beetle broods. The green-attacked trees are ``sanitized'' (i.e., burned or chipped) to prevent the proliferation and spread of mountain pine beetle \citep{government2016mountain}.

\subsection{Data}

\begin{figure}[!ht]
\centering
\includegraphics[scale = 1]{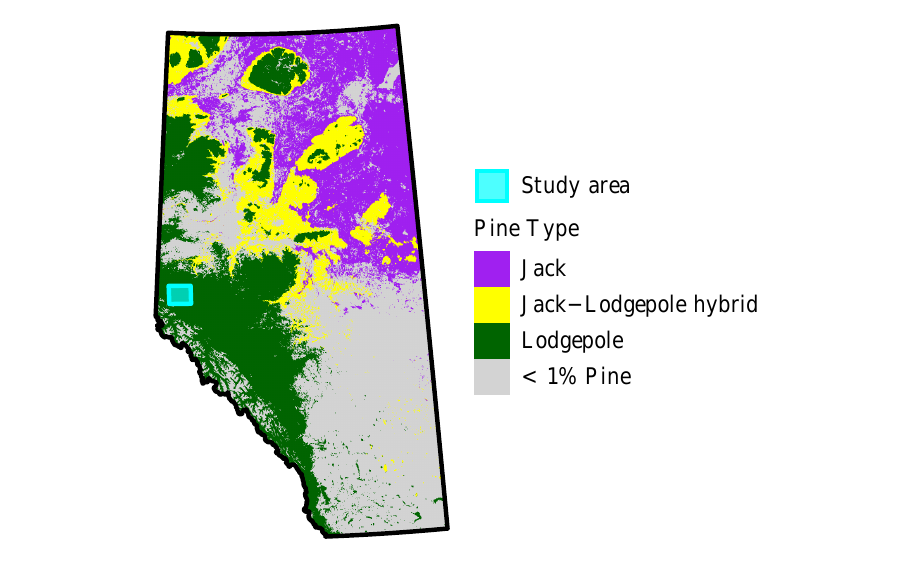}
\caption{Location of our study area in Alberta. Our study area (cyan) is an approximately 2,500 $\text{km}^2$ block of high-biomass lodgepole pine, and was surveyed annually from 2006--2021. Pine species data comes from \citet{cullingham2012characterizing}. The grey pixels are areas where pines constitute less than 1\% of total live aboveground biomass; data from \citet{beaudoin2014mapping}.}
\label{fig:study_area_smallest}
\end{figure}

Our paper utilizes data from the Alberta Heli-GPS and ground surveys, from 2005--2020. We restricted our main analysis to an approximately 2500 $\text{km}^2$ patch of lodgepole pine forest in western Alberta (Fig.~\ref{fig:study_area_smallest}). This study area was selected because it contains suitable mountain pine beetle habitat (i.e., high-biomass forest, moderate temperatures), is relatively homogeneous (i.e., low topological complexity, contiguous forest), and was surveyed every year. For robustness, we replicated our analysis on a second 2500 $\text{km}^2$ patch that is approximately 50 km east of the original study area; the results are qualitatively identical, and are thus relegated to Appendix \ref{app:robust}.

\subsection{Data preparation}

The raw heli-GPS and ground survey data contain GPS coordinates and the number of trees (red-topped or controlled trees respectively). We rasterized this point data into 30 x 30-meter pixels. Rasterizing with a higher resolution would likely offer no benefits, seeing that location data may have positional errors up to $\pm$ 30 meters. 

We use $I_t(x)$ to denote the number of infested trees in pixel $x$, in year $t$. The number of infestations is calculated as 
\begin{equation}
I_t(x) = c_{t}(x) + r_{t+1}(x),
\end{equation}
where $c_{t}(x)$ is the number of green-attack trees that were located and ``controlled'' (i.e., burned or chipped) in the focal year, and $r_{t+1}(x)$ is the number of red-topped trees observed in the following year. Our overarching goal is to predict the positions of next year's infestations, given this year's infestations. However, $I_t$ is not a suitable model input --- it contains green-attack trees that are controlled, and thus cannot contribute to future infestation. Therefore, we define a slightly modified input variable, $I_{t}^{*}(x) = I_t(x) - c_t(x)$.

\subsection{Model}

\begin{figure}[H]
\centering
\makebox[\textwidth]{\includegraphics[scale = 1]{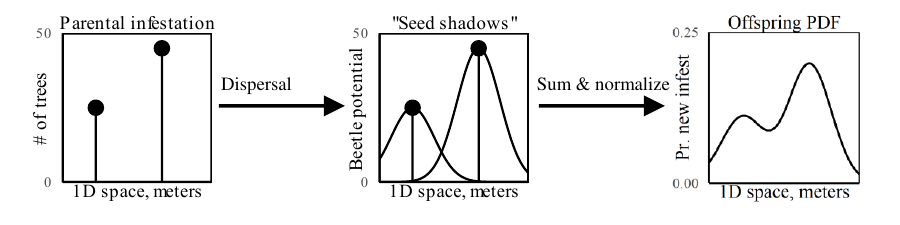}}
\caption{Graphical representation of our redistribution models in one dimension. A dispersal model is applied to each parental infestation, creating surfaces that represent the potential of offspring infestations stemming from parental infestation. These are called ``seed shadows'' in the plant ecology literature. To create a probability density function for new infestations, we sum and normalize these seed shadows.}
\label{fig:model_conceptual_fig}
\end{figure}

Conceptually, our models of mountain pine beetle dispersal are simple (Fig. \ref{fig:model_conceptual_fig}). A dispersal kernel predicts the spatial distribution of infestations in the focal year (\textit{offspring infestations}), stemming from each infestation in the prior year (\textit{parental infestations}). All of these individualized distributions are combined to create a landscape-scale dispersal distribution, and then normalized to create a probability distribution for the offspring infestations. Unlike heuristic methods that calculate distances between offspring infestations and their closest parental source, the model-based approach offers a key advantage by averaging over multiple possible origins for each offspring infestation.

We examined a variety of different sub-models for dispersal, summarized in Table \ref{tab:all_kernels}. The general mathematical form of dispersal kernel is $D(r = dist(x,y))$. The function $D$ maps the destination coordinate $y$ and the source coordinate $x$ to a probability density. Here, each dispersal kernel is assumed to be radially symmetric, and thus each kernel can be parameterized as a function of the Euclidean distance between coordinates, $r = dist(x,y)$. Although we will use the term ``dispersal kernel'' for its familiarity, the jargon ``redistribution kernel'' is technically more accurate, since we are not modelling beetle dispersal \textit{per se}, but rather the spatial relationship between infestations.

There are three classes of dispersal kernels: thin-tailed, fat-tailed, and mixtures of thin-tailed distributions. The moniker ``thin-tail'' describes any distribution whose tails decay exponentially (or faster) in the limit of large $r$. The Gaussian distribution is a prime example of this category, exhibiting the fastest decay rate among all the distributions we examined. By contrast, fat-tailed distributions decay slower than exponentially. The Pareto distribution is a prime example of this category, with tails so fat the variance of the distribution doesn't exist when the scale parameter $\nu < 2$.  Mixture distributions, which are weighted averages of two thin-tailed distributions, technically remain thin-tailed due to their asymptotic behavior, but can simulate fat-tailed distributions over intermediate distances by blending distributions with various distance scales. The flexible nature of mixtures is illustrated by the radially symmetric Student's $t$-distribution, also referred to as the ``2Dt dispersal kernel''. This properly fat-tailed distribution can be derived as an infinite mixture of Gaussian distributions whose variances are prescribed to follow an inverse-gamma-distribution~\citep{lewis_mathematics_2016}. 

\renewcommand{\arraystretch}{2}
\begin{table}[H]
\caption{All of the dispersal kernels that were fit to the Alberta Heli-GPS data, along with their tail-type classification, functional form, and mechanistic interpretation. The Whittle–Matérn–Yasuda kernel (abbr.~WMY) contains the function $K_\kappa$ denoting the $\kappa^\text{th}$ order modified Bessel function of the second kind. The parameter $\nu$ is constrained to be greater than 1; otherwise, the normalizing constant of the distribution does not exist. This table was inspired by Table 1 in \citet{koch_unifying_2020}.}
\label{tab:all_kernels}
\centering
\begingroup\fontsize{6pt}{7pt}\selectfont
\noindent\makebox[\textwidth]{%
\begin{tabular}{lll p{0.5\textwidth}}
  \toprule
Kernel name & Tail type & $D(r) \propto$ & Mechanistic derivation \\ 
  \midrule
Pareto & Fat & $(r + \rho)^{-(1+\nu)}$  & Levy random walk; successive dispersal steps from a mixture of distributions with disparate characteristic length scales  \citep{benhamou_how_2007} \\
  Student's $t$ & Fat & $\left(1 + \frac{1}{\nu}\left(\frac{r}{\rho}\right)^2\right)^{-\frac{\nu + 1}{2}}$ & 2D Fickian diffusion with inverse-gamma-distributed settling times \citep[][pg. 167]{lewis_mathematics_2016} \\ 
  Bessel mixture & Mixture of Thin & $\theta \, K_{0}(\frac{r}{\rho_1}) + (1 - \theta)K_{0}(\frac{r}{\rho_2})$ & 2D Fickian diffusion with two different motilities and/or two different settling rates.  \\ 
  Laplace mixture & Mixture of Thin & $\theta\exp\left(-\frac{r}{\rho_1}\right) + (1 - \theta)\exp\left(-\frac{r}{\rho_2}\right)$ & 2D Fickian diffusion with two different motilities and instantaneous settling. \\ 
  Gaussian mixture & Mixture of Thin & $\theta\exp\left(-\left(\frac{r}{\rho_1}\right)^2\right) + (1 - \theta)\exp\left(-\left(\frac{r}{\rho_2}\right)^2\right)$ & 2D Fickian diffusion with two different motilities and a single settling time, or a single motility constant and two different settling times.\\ 
  WMY & Thin & $\left(\frac{r}{\rho}\right)^{\kappa}  K_{\kappa}(\frac{r}{\rho})$ & 2D Fickian diffusion with gamma-distributed stopping times \textit{or} 2D fractal diffusion with constant settling rate \citep{koch_unifying_2020} \\ 
  Bessel & Thin & $K_{0}(\frac{r}{\rho})$ &  2D Fickian diffusion with constant settling rate \citep{broadbent_random_1953} \\ 
  Laplace & Thin & $\exp\left(-\frac{r}{\rho}\right)$ & Turbulent diffusion then instantaneous settling \citep{joseph_uber_1958} \textit{or} a bout of 2D Fickian diffusion followed by a bout of fractal diffusion, then instantaneous settling \citep{koch_unifying_2020} \\ 
  Gaussian & Thin & $\exp\left(-\left(\frac{r}{\rho}\right)^2\right)$ & 2D Fickian diffusion then instantaneous settling \citep{skellam_random_1951} \\ 
   \bottomrule
\end{tabular}
}
\endgroup
\end{table}
\renewcommand{\arraystretch}{1}


The likelihood function is computed by first convolving the map of the parental infestations, $I_{t-1}^*(x)$, with the dispersal kernel:
\begin{equation} \label{eq:convolve}
B_t(y) = \sum_{x} I_{t-1}^*(x) D\left(\text{dist}(y, x)\right) \left( \Delta x \right)^2,
\end{equation}
with $\Delta x = 0.03 \text{km}$ as the spatial resolution. The result is the \textit{beetle potential} $B_t$, which represents the  average number of beetles arriving at each location. Note that \eqref{eq:convolve} is the discretization of a continuous-space convolution; thus, $\text{dist}(y, x)$ is the distance between the centers of two 30-meter pixels, regardless of the spatial distribution of infestations within those pixels. 

The beetle potential is re-scaled to produce a likelihood surface:
\begin{equation}
\pi_t(x) = \frac{B_t(x)}{\sum_y I_{t-1}^*(y)}
\end{equation}

While the convolution is performed for all infestations within the approximately 50x50 km study area, the likelihood is only calculated using the new infestations that were further than 10km from the boundary of the study area. This buffer zone circumvents statistical edge effects, ensuring that each modeled infestation is allowed to receive virtual beetles from all directions. Each offspring infestation is treated as an i.i.d. event, and thus the log likelihood in year $t$ is
\begin{equation}
\mathcal{L}_t = \sum_{x:I_t(x) > 0} I_t(x) \times \log(\pi_t(x)). 
\end{equation}
The total log likelihood is simply the sum over years, from 2009 to 2019: $\mathcal{L} = \sum_{t=1}^{11}  \mathcal{L}_t$. While the presence of MPB in Alberta dates back to 2005, the years 2009--2019 constitute the main phase of heightened MPB presence (Fig. \ref{fig:infest_time_series}).

All dispersal kernels were fit with maximum likelihood estimation. More specifically, multi-parameter kernels were fit with the Nelder-Mead algorithm, whereas the single-parameter kernels (i.e. the Laplace, Gaussian, and Bessel) were optimized using high-resolution likelihood profiles (Fig.~\ref{fig:ll_profiles}). All analyses were performed in R \citep{r_core_team_r_2022}


\subsection{Model validation}

The dispersal models were validated and compared using a holistic approach that combined metrics of relative model fit (e.g.~the log likelihood), absolute model fit (e.g.~forecast correlation, true detection rate), and the visual agreement between model predictions and salient features of the data. Notably, we calculated the distance between offspring infestations and the nearest parental infestation. This nearest neighbor analysis produces a lower bound for model-based estimates of dispersal distance, indicates the general shape of the marginal dispersal kernel, and helps us roughly classify observations as short or long-distance dispersal events. This classification enables us to assess models more effectively by examining the log likelihood of observations within defined categories of dispersal distance. 

To explore the long-term implications of the different dispersal kernels, we simulated the spread of MPB across Alberta. Each simulation started with the same initial conditions, the observed infestations in 2005. Then, for each successive year, the simulated infestations determine the beetle-potential surface and corresponding probability mass function, from which the new infestations are simulated. The number of simulated infestations is equal to the actual number of observed infestations in the focal year; this methodology accounts for the fact that some years (particularly the 2008--2009 transition) exhibit variable reproductive rates, potentially because of fluctuating climatic factors.  The simulation was constrained to areas where lodgepole pine is the dominant pine species and where pine biomass constitutes more than 1\% of total live aboveground biomass.

\section{Results}

\begin{figure}[H]
\centering
\makebox[\textwidth]{\includegraphics[scale = 1]{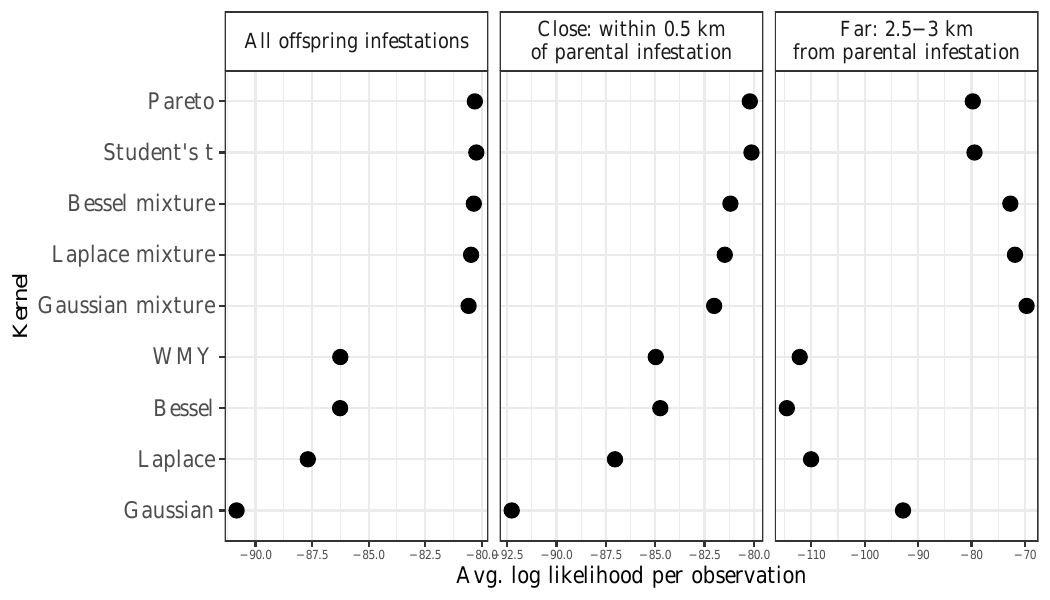}}
\caption{Model fit for observations that are various distances from the nearest parental infestations. Observations that were within 0.5 kilometers of the nearest parental infestation (representing short-to-medium distance dispersal) were best modeled by the fat-tailed distributions, the Pareto and Student's $t$-distributions. Observations that were within 2.5--3 km from the nearest parental observation were best modeled by the mixture distributions.}
\label{fig:ll_by_dist}
\end{figure}

Across all forms of evidence, the most effective models featured fat-tailed dispersal kernels. There were larger differences between the three categories of dispersal kernels --- thin-tailed, fat-tailed, and mixture --- than there were within each category (Fig.~\ref{fig:ll_by_dist}, leftmost panel). The differences between fat-tailed and mixture-based dispersal kernels are subtle, whereas models with thin-tailed dispersal kernels consistently performed worse by all measures of quality.

\begin{figure}[H]
\centering
\makebox[\textwidth]{\includegraphics[scale = 1]{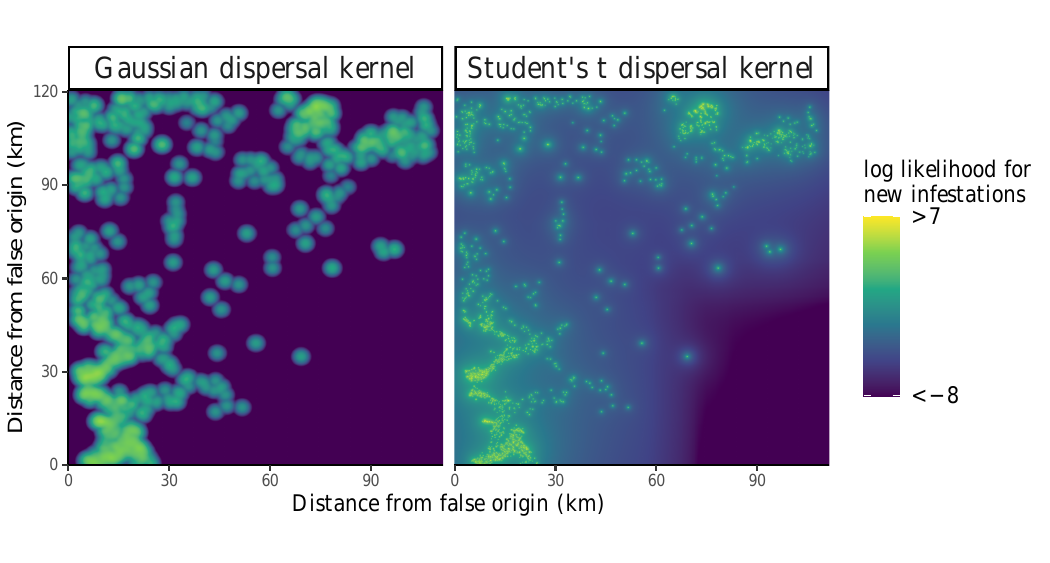}}
\caption{In order to account for medium-to-long distance dispersal events, thin-tailed redistribution kernels (here, the radially-symmetric Gaussian) predict that offspring infestations will be several kilometers from parental infestations. The fat-tailed dispersal kernels (here, the radially-symmetric Student's $t$) predominantly predict short-distance dispersal, along with a small number of long-distance dispersal events. The maps display model predictions of beetle potential from trees that were infested in 2006. The origin has coordinates longitude \& latitude coordinates -119.7269,53.60261.}
\label{fig:bp_2D}
\end{figure}

Thin-tailed dispersal kernels, like the familiar Gaussian and Laplace kernels, consistently produced low log likelihoods, predictive correlations, and true positive rates for new infestations (Table \ref{tab:redist_stats1}). There is a huge log likelihood penalty for dispersal kernels that predict a near-zero infestation probability for a pixel in which an infestation does occur; to avoid this penalty, the thin-tailed dispersal kernels predict that most dispersal is medium-distance dispersal. This phenomenon is most evident in the case of the Gaussian dispersal kernel, which has the fastest-decaying tails of any distribution considered here, and consequently predicts a median dispersal distance of 1.1 km (Table \ref{tab:redist_stats1}). This is probably wrong, since the majority of experiments and models --- including our other dispersal kernels --- imply that most beetles disperse much shorter distances. The general pattern, wherein thin-tailed kernels predict a predominance of medium-distance dispersal, is readily visualized with a 2D log-likelihood surface (Fig.~\ref{fig:bp_2D}).

\begin{table}[H]
\caption{Summary statistics of model fit and redistribution distances for the redistribution models, fit to data from study area \#1. The abbreviation TPR stands for \textit{True Positive Rate}; $r$ is the correlation between the logarithm of the observed number of infestations and the logarithm of the expected number of infestations; and the last 4 columns refer to the distances between new infestations and their parental infestations. To compute the TPR, we identify a ``positive'' as one or more infestations within a 30 x 30-meter pixel, and we predict a positive when the probability of one or more infestations is greater than 1/2. This probability is computed as the complement of the Bernoulli probability of observing zero new infestations, given a number of trials equal to the total infestations observed in the focal year, and the per trial probability is derived from the redistribution model. The expected number of infestations (used in the calculation of $r$) is simply the per-trial probability of an infestation, multiplied by the total number of infestations in the focal year.}
\label{tab:redist_stats1}
\centering
\begingroup\fontsize{8pt}{9pt}\selectfont
\noindent\makebox[\textwidth]{%
\begin{tabular}{llllllll}
  \toprule
Kernel name & Log likelihood & TPR & $r$ (log scale) & mean dist. & median dist. & 75\% dist. & 95\% dist. \\ 
  \midrule
Pareto & $-7.494 \cdot 10^{5}$ & 0.009 & 0.327 & 0.957 & 0.080 & 0.320 & 3.950 \\ 
  Student's $t$ & $-7.488 \cdot 10^{5}$ & 0.015 & 0.330 & 1.059 & 0.060 & 0.270 & 4.670 \\ 
  Bessel mixture & $-7.498 \cdot 10^{5}$ & 0.004 & 0.343 & 2.571 & 0.070 & 3.440 & 12.940 \\ 
  Laplace mixture & $-7.510 \cdot 10^{5}$ & 0.004 & 0.346 & 3.043 & 0.070 & 4.850 & 13.920 \\ 
  Gaussian mixture & $-7.520 \cdot 10^{5}$ & 0.013 & 0.348 & 4.695 & 1.730 & 8.730 & 15.690 \\ 
  WMY & $-8.049 \cdot 10^{5}$ & 0.000 & 0.170 & 0.355 & 0.280 & 0.480 & 0.900 \\ 
  Bessel & $-8.050 \cdot 10^{5}$ & 0.000 & 0.168 & 0.338 & 0.270 & 0.460 & 0.860 \\ 
  Laplace & $-8.183 \cdot 10^{5}$ & 0.000 & 0.148 & 0.429 & 0.360 & 0.580 & 1.020 \\ 
  Gaussian & $-8.477 \cdot 10^{5}$ & 0.000 & 0.096 & 0.982 & 0.920 & 1.300 & 1.920 \\ 
   \bottomrule
\end{tabular}
}
\endgroup
\end{table}

Fat-tailed kernels produce higher log likelihoods than mixture-based kernels, but the difference is slight, thus suggesting that the two types of dispersal kernels be differentiated based on additional features. Both predict a majority of short-distance dispersal and a minority of long-distance dispersal, but the mixture models are more extreme. To see this, consider the Laplace kernel with density $\exp(-r/\rho)$, which has the nice property of the mean distance equaling $2\times \rho$ km. The Laplace mixture model predicts that a $\theta= 0.62$ proportion of beetles engage in short-distance dispersal with a mean distance of 42 meters, while the remaining beetles disperse with a mean distance of 7.8 km (Table \ref{tab:redist_pars1}).

\renewcommand{\arraystretch}{2}
\begin{table}[H]
\caption{Parameter values for the redistribution models, fit to data from study area \#1. The parameters $\rho$, $\rho_1$, and $\rho_2$ have units of kilometers; the remaining parameters are dimensionless.} 
\label{tab:redist_pars1}
\centering
\begingroup\fontsize{6pt}{7pt}\selectfont
\noindent\makebox[\textwidth]{%
\begin{tabular}{llll}
  \toprule
Kernel name & Tail type & $D(r) \propto$ & Max likelihood estimate \\ 
  \midrule
Pareto & Fat-tail & $\left(r + \rho\right)^{-(1+\nu)}$ & $\rho = 1.35 \cdot 10^{-2}, \;\nu = 1.57$ \\ 
  Student's $t$ & Fat-tail & $\left(1 + \frac{1}{\nu}\left(\frac{r}{\rho}\right)^2\right)^{-\frac{\nu + 1}{2}}$ & $\rho = 1.18 \cdot 10^{-2}, \;\nu = 1.45$ \\ 
  Bessel mixture & Mixture of Thin & $\theta \, K_{0}(\frac{r}{\rho_1}) + (1 - \theta)K_{0}(\frac{r}{\rho_2})$ & $\rho_1 = 3.20 \cdot 10^{-2}, \;\rho_2 = 4.62, \;\theta = 6.56 \cdot 10^{-1}$ \\ 
  Laplace mixture & Mixture of Thin & $\theta\exp\left(-\frac{r}{\rho_1}\right) + (1 - \theta)\exp\left(-\frac{r}{\rho_2}\right)$ & $\rho_1 = 2.12 \cdot 10^{-2}, \;\rho_2 = 3.91, \;\theta = 6.18 \cdot 10^{-1}$ \\ 
  Gaussian mixture & Mixture of Thin & $\theta\exp\left(-\left(\frac{r}{\rho_1}\right)^2\right) + (1 - \theta)\exp\left(-\left(\frac{r}{\rho_2}\right)^2\right)$ & $\rho_1 = 2.71 \cdot 10^{-2}, \;\rho_2 = 1.03 \cdot 10^{1}, \;\theta = 4.91 \cdot 10^{-1}$ \\ 
  WMY & Thin tail & $\left(\frac{r}{\rho}\right)^{\kappa} K_{\kappa}(\frac{r}{\rho})$ & $\rho = 2.26 \cdot 10^{-1}, \;\kappa = 3.95 \cdot 10^{-9}$ \\ 
  Bessel & Thin tail & $K_{0}(\frac{r}{\rho})$ & $\rho = 2.15 \cdot 10^{-1}$ \\ 
  Laplace & Thin tail & $\exp\left(-\frac{r}{\rho}\right)$ & $\rho = 2.15 \cdot 10^{-1}$ \\ 
  Gaussian & Thin tail & $\exp\left(-\left(\frac{r}{\rho}\right)^2\right)$ & $\rho = 1.11$ \\ 
   \bottomrule
\end{tabular}
}
\endgroup
\end{table}
\renewcommand{\arraystretch}{1}



\begin{figure}[H]
\centering
\includegraphics[scale = 1]{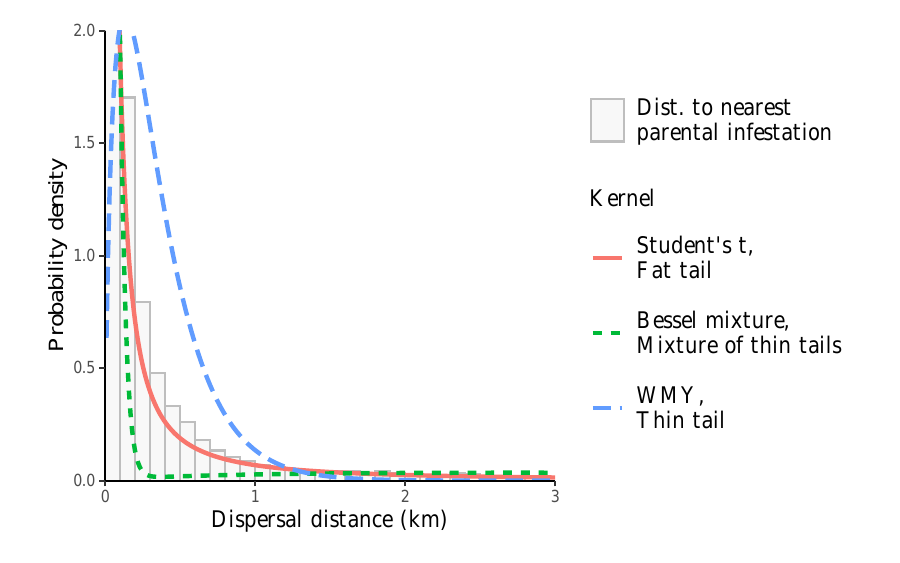}
\caption{Predictions vs. observations for various dispersal kernels. The histogram gives the frequency of distances between infestations and the closest ``parental infestation'' from the previous year. The predictions come from the marginal dispersal kernels $D_r(r)$, i.e., $\int_{0}^{2\pi} D(r) r d\theta = 2\pi\,r\,D(r)$, where $r$ is the distance in kilometers, $\theta$ is the angle in radians, and $D(r)$ is the full density (see Table \ref{tab:all_kernels}). For each class of distributional tail --- thin, mixed, and fat --- we display the model with the highest log likelihood.} 
\label{fig:marginal_kernels}
\end{figure}

There are several reasons for treating fat-tailed dispersal kernels as the default choice when modeling MPB. 1) Fat-tailed kernels provide a much better fit to the non-model dispersal kernel, computed with the minimum distance between parental and offspring infestations (Fig.~\ref{fig:marginal_kernels}). 2) Mixture models are less parsimonious, with $\geq 3$ parameters. 3) According to the distance-stratified log likelihoods (Fig.~\ref{fig:ll_by_dist}), mixture-based kernels are better at predicting offspring infestations that are far away from any parental infestation. However, fat-tailed kernels are better at predicting nearby offspring infestations, which constitute the vast majority of new infestations.

\section{Discussion} \label{sec:discussion}

\begin{figure}[H]
\centering
\makebox[\textwidth]{\includegraphics[scale = 1]{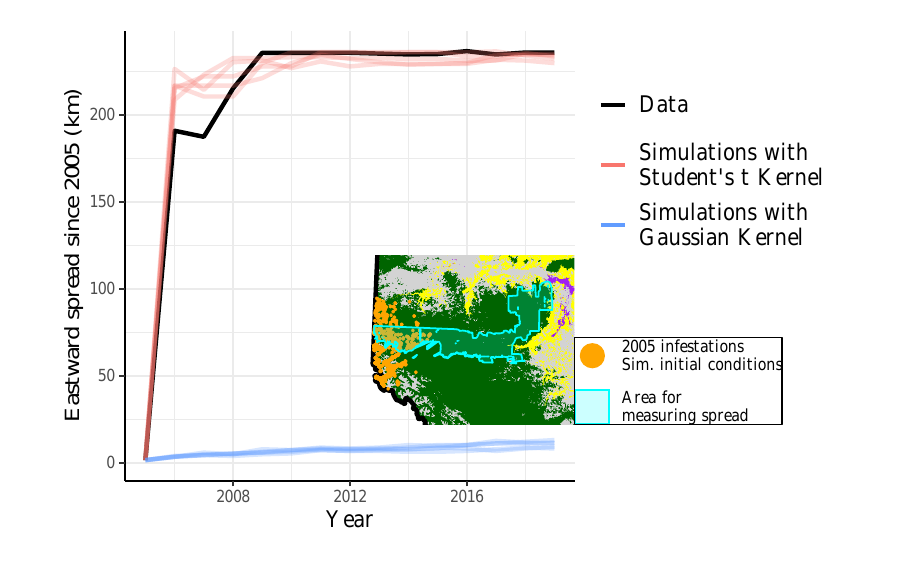}}
\caption{Fat-tailed redistribution kernels can explain the spread of MPB across Alberta. Using the 2005 infestations as the initial condition, we simulated the spatial spread of the MPB across the lodgepole pine zone of western Alberta (the dark green area in the inset plot). To summarize the spatial extent of infestations, we calculated the maximum easting coordinate for all infestations within the cyan polygon (see inset plot). This polygon is the intersection of Heli-GPS survey areas from 2005--2019.}
\label{fig:sim_spread_w_inset}
\end{figure}

Fat-tailed dispersal kernels, particularly the Student's $t$-distribution characterized by degrees of freedom $\nu = 1.45$ and scale parameter $\rho = 0.0118$, emerge as the most effective way to model MPB dispersal. The Student's $t$ kernel yields dispersal distances with a median, mean, and $95^{\text{th}}$ percentile of 0.080, 1.06, and 4.67 kilometers respectively. Unlike mixture-based dispersal kernels, the Student's $t$-distribution faces no numerical model-fitting problems, only has 2 parameters, and successfully re-creates empirical patterns of distances between parental and offspring infestations (Fig.~\ref{fig:marginal_kernels}). Although such a simple model cannot hope to capture \textit{all} aspects of MPB dispersal, the Student's $t$-distribution serves as a useful first-order approximation for both researchers and conservation practitioners. Indeed, our stochastic simulations reveal that a simple fat-tailed dispersal kernel outperforms thin-tailed kernels in reconstructing the mountain pine beetle's eastward expansion (Fig. \ref{fig:sim_spread_w_inset}).

The Student's $t$-distribution has a clear interpretation in light of MPB biology: the Student's $t$-distribution can be derived as an infinite mixture of Fickian diffusion processes with a gamma-distribution for stopping times or, equivalently, for the diffusion coefficients. It is this mixture that represents the variety of MPB dispersal strategies. Three types of dispersal behaviors roughly correspond to three classes of dispersal distances: Dispersal behaviors can be categorized into three types: short-range (typically < 100 m), medium-range (up to 5 km), and long-range (above canopy, potentially up to 300 km). These distance ranges are approximate and illustrate orders of magnitude associated with each category. Short-range dispersal predominantly occurs under the canopy, with a large majority of beetles attracted to nearby plumes of aggregation pheromones \citep{safranyik_dispersal_1992, robertson_mountain_2007, dodds_sampling_2002}. Medium-range dispersal involves a minority of beetles traveling up to 5 km to initiate pioneer attacks, resulting in semi-regular patterns of so-called spot infestations \citep{strohm_pattern_2013}. Proposed explanations for these pioneer attacks include evolutionary bet-hedging strategies \citep{raffa_mixed_2001, kautz_dispersal_2016} and the absence of guiding pheromones for early-emerging beetles. Long-range dispersal, observed in just 0--2.5\% of beetles \citep{robertson_spatialtemporal_2009, safranyik_dispersal_1992}, involves flight above the canopy, where beetles may be swept up by atmospheric winds and carried up to 300 km \citep{hiratsuka_forest_1982, cerezke_mountain_1989, jackson_radar_2008}. Fat-tailed dispersal kernels effectively interpolate between the three dispersal behaviors, thus capturing a wide range of dispersal distances.

Consistent with an inverse-gamma distribution of settling times, a recent flight-mill experiment shows that many beetles fly short distances, while few fly long distances (range 0--25 km; \citealp{evenden_factors_2014}). The basis for this intraspecific variation is unclear. Larger beetles, which tend to emerge from larger trees \citep{graf2012association}, fly further on average. However, their flight distance is also more variable with different individuals flying short or long distances \citep{shegelski_morphological_2019}. This polyphenism is thought to be an evolutionarily stable bet-hedging strategy \citep{kautz_dispersal_2016, jones2020mechanisms}, where the short-distance dispersers are more likely to find host trees and conspecifics with which to perform mass attacks, and the risky longer-distance dispersers may enjoy the fitness advantages of arriving early to a large tree (\citealp{raffa_mixed_2001} and sources therein).

Rare, long-distance ``jackpot-dispersal'' events, as described by the tails of fat-tailed dispersal kernels, are key for rapid spread. For example, in the case of MPB there is evidence that their range expansion across Alberta jumped eastward 250 km in a single year \citep{cooke_predicting_2017}. These events can also yield a characteristic signature pattern of patchy spread on the landscape. Such patterns are most easily seen from Monte Carlo simulations of stochastic invasion processes that have a mixture of rare long- and common short-distance dispersal distances ~\citep{lewis2000modeling}. The resulting spatial process, sometimes referred to as ``stratified diffusion''~\citet{shigesada_modeling_1995}, exhibits a characteristic patchy pattern of spread, with rare long-distance dispersal events creating new invasion beachheads at a distance, followed by common short-distance dispersal coupled to growth dynamics generating growth of patches on a local scale.  While a full mathematical analysis of such processes typically involves spatial moments of the associated point process of invasion~\citep{lewis2000modeling}, evidence of patchy spread in the spread of MPB can be seen visually from Figure \ref{fig:bp_2D}.  

The models we use are intentionally simple compared to the complex biology of MPB, but this simplicity is deliberate and serves a clear purpose. Beetle dispersal is influenced by many factors: at small spatial scales, it involves the phenology of beetle emergence, the production and diffusion of aggregation and anti-aggregation pheromones, forest micro-climates, and wind patterns \citep{amman_silvicultural_1998, biesinger_direct_2000}; at larger spatial scales, dispersal involves temperature patterns, wind-storms, and the availability of host trees \citep{safranyik_forest_2006, powell_phenology_2014}. We contend that incorporating all these factors into the model is unnecessary and potentially counterproductive because 1) the specificity required for a detailed model would likely reduce its applicability to different areas or time periods, making it less useful for practitioners and scientists seeking to apply our results more broadly; and 2) a simple model with clear assumptions minimizes the risk of significant errors. In other words, we can be assured that dispersal distance estimates are not merely artifacts of model structure. 

Previous research has provided a large range of dispersal estimates (recall Table \ref{tab.data}). In particular, the work of \citet{heavilin_novel_2008} resulted in an implausibly small estimate of median dispersal distance: about 10 meters. This result can be attributed to a least-square model fitting procedure, which inadequately penalizes the prediction of zero new infestations in areas where infestations occur. Our model addresses these limitations by introducing a probabilistic landscape for new infestations, such that assigning a near-zero probability to an actual infestation incurs a near-infinite penalty. On the other side of the spectrum, the model of \citet{koch_signature_2021} implies an implausibly large estimate of the median dispersal distance: about 17 km. This can be explained by the two-part structure of their model, which includes a dispersal sub-model and a pine susceptibility sub-model, where susceptibility is an increasing function of last year's infestations. Therefore, the short distances between successive years' infestations --- normally attributed to short dispersal distance --- are accounted for in the susceptibility sub-model, leaving the dispersal parameters relatively unconstrained. Our approach allows for a more accurate estimation of dispersal distances by focusing on the redistribution of infestations rather than individual beetles, and by taking into account inter-annual variation in beetle productivity through the predetermined number of new infestations.

The mathematical theory for spreading populations is based on models coupling dispersal and population growth dynamics; it generally assumes no significant Allee effect (i.e., positive density dependence) in the population growth dynamics~\citep{lewis_mathematics_2016}. We take this approach in our models despite MPB's well-known Allee effect \citep{safranyik1975interpretation, raffa_role_1983, boone_efficacy_2011}, which occurs because MPB must mass-attack trees to deplete the trees' resin defenses. One justification is that MPB's Allee threshold --- the population density at which per capita growth rates are negative --- is small and varies with environmental conditions \citep{cooke2024deduction}. Another justification is that wind currents can carry large groups of beetles to the same destination, such that long-distance dispersal can be modeled as the movement of groups large enough to surpass the Allee threshold. 

One advantage of using a dispersal kernel to model MPB is that it is relatively simple to understand and easily applied. The trade-off is that we do not capture all biological aspects of MPB dispersal. In particular, we do not include clustering mechanisms for dispersal --- for long-distance dispersal it would be more accurate to assume beetles travel \textit{en masse}, carried by wind currents --- nor do we include clustering of beetles at small spatial scales to represent chemical signaling. This leads to a spatial distribution of simulated infestations that is unrealistically patchy (Fig. \ref{fig:sim_student_t}). Additionally, our simulations (which generate Fig. \ref{fig:sim_spread_w_inset}) do not incorporate the effects of host tree depletion and assume pine density is constant, which is roughly true over short time scales but may not be true in heavily infested areas over entire outbreaks. 
Finally, beetles may disperse without necessarily infesting trees, potentially leading to underestimation of dispersal distances (as in Table \ref{tab:redist_stats1}). This bias arises from the Allee effect: beetles dispersing farther from infestation clusters have fewer companions for coordinating mass attacks, reducing their likelihood of successful attack and subsequent detection by Heli-GPS surveys. However, for reasons discussed in the previous paragraph (i.e., the clumped nature of dispersal and the apparent weakness of the Allee effect) we expect this bias to be minimal.

Our conclusions may also be particular to the study area or the years of study. We address these concerns about generalizability by considering an additional area 50 km east of the initial study area, as outlined in methods, and find qualitatively similar results (see Appendix~\ref{app:robust}). We may additionally expect inter-annual variation in beetle dispersal given that dispersal distance may change with fluctuating environmental factors  \citep{chen_climatic_2017,mccambridge_temperature_1971,powell_phenology_2014,wijerathna_effect_2020}; but note that \citet{carroll_assessing_2017} find similar distances between years. We explored inter-annual variability by applying a Student's t-dispersal kernel to each year of data independently (Appendix \ref{app:interannual}) though a comprehensive analysis remains beyond this paper's scope. The median dispersal distance fluctuates with a right-skewed distribution: most years show low distances (less than 100 meters), while a few years exhibit much larger distances (several hundred meters).

In conclusion, we find that relatively simple fat-tailed kernels can accurately characterize MPB dispersal. These kernels can describe both the short and long-distance dispersal modes of mountain pine beetle, which are both important at the landscape scale. In particular, the long-distance dispersal events captured by fat-tailed kernels are key for describing range expansion. We expect that this result applies for other species undergoing range expansions, and points to the importance of simple but accurate dispersal kernels for understanding and modeling these processes.

\pagebreak

\section{Acknowledgements}
The authors would like to thank Xiaoqi Xie and K{'e}van Rastello for feedback. Funding for this research has been provided through grants to the TRIA-FoR Project to ML from Genome Canada (Project No. 18202) and the Government of Alberta through Genome Alberta (Grant No.~L20TF), with contributions from the University of Alberta and fRI Research (Project No. U22004). This work was supported by Mitacs through the Mitacs Accelerate Program, in partnership with fRI Research. MB acknowledges the support of the Natural Sciences and Engineering Research Council of Canada (NSERC), [PDF – 568176 - 2022].

\section{Author contributions}
All authors conceived of the project; Evan C. Johnson performed the analysis; Evan C. Johnson and Micah Brush wrote the first draft; All authors contributed critically to the drafts and gave final approval for publication.

\section{Data sources} \label{Data and code availability statement}
Code and data will be made available on Zenodo (\url{TBD}).

\newpage

\begin{appendices}
\counterwithin{figure}{section}
\counterwithin{table}{section}

\section{Robustness check: Study area \#2} \label{app:robust}

\begin{figure}[H]
\centering
\makebox[\textwidth]{\includegraphics[scale = 0.7]{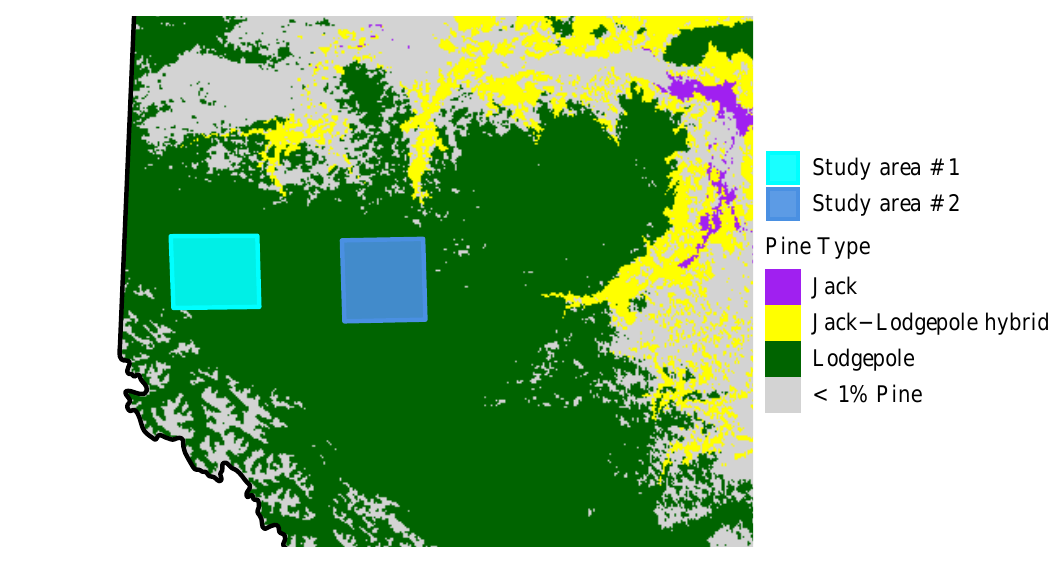}}
\caption{The location and extent of both study areas in Alberta.}
\label{fig:study_area_smallest_both}
\end{figure}

The nature of MPB dispersal may vary across space, possibly due to un-modelled factors such as MPB population density, pine density, wind patterns, and annual temperature patterns. As reassurance that our dispersal models will produce reasonable results, even if applied outside of our focal study area, we re-ran our analysis in a second study area. This area is located approximately 50 km east of study area \#1 (Fig.~\ref{fig:study_area_smallest_both}) and has approximate dimensions of 50x50 km.

In study area \#2, the estimated median and mean dispersal distances for the Student's $t$ kernel are 0.060 and 0.988 km respectively (Table \ref{tab:redist_stats2}). These numbers are extremely similar to the estimates from study area \#1 (Table \ref{tab:redist_stats1}). In study area \#2, fat-tailed dispersal kernels are superior to mixture-based dispersal kernels with respect to the log likelihood; the true positive rate of predicting new infestations; and the correlation between log-scale predictions and infestations. For completeness, parameter estimates for study area \#2 are provided in Table \ref{tab:redist_pars2}.

\begin{table}[H]
\caption{Summary statistics of model fit and redistribution distances for the redistribution models, fit to data from study area \#2. The abbreviation TPR stands for \textit{True Positive Rate}; $r$ is the correlation between the logarithm of the observed number of infestations and the logarithm of the expected number of infestations; and the last 4 columns refer to the distances between new infestations and their parental infestations.}
\label{tab:redist_stats2}
\centering
\begingroup\fontsize{8pt}{9pt}\selectfont
\noindent\makebox[\textwidth]{%
\begin{tabular}{llllllll}
  \toprule
Kernel name & Log likelihood & TPR & $r$ (log scale) & mean dist. & median dist. & 75\% dist. & 95\% dist. \\ 
  \midrule
Pareto & $-1.491 \cdot 10^{6}$ & 0.094 & 0.299 & 0.847 & 0.080 & 0.290 & 3.310 \\ 
  Student's $t$ & $-1.491 \cdot 10^{6}$ & 0.108 & 0.295 & 0.988 & 0.060 & 0.250 & 4.190 \\ 
  Bessel mixture & $-1.499 \cdot 10^{6}$ & 0.082 & 0.298 & 0.542 & 0.070 & 0.790 & 2.360 \\ 
  Laplace mixture & $-1.504 \cdot 10^{6}$ & 0.097 & 0.289 & 0.495 & 0.080 & 0.800 & 1.900 \\ 
  Gaussian mixture & $-1.518 \cdot 10^{6}$ & 0.106 & 0.274 & 0.710 & 0.500 & 1.260 & 2.170 \\ 
  WMY & $-1.582 \cdot 10^{6}$ & 0.000 & 0.241 & 0.307 & 0.250 & 0.420 & 0.780 \\ 
  Bessel & $-1.582 \cdot 10^{6}$ & 0.000 & 0.246 & 0.338 & 0.270 & 0.460 & 0.860 \\ 
  Laplace & $-1.605 \cdot 10^{6}$ & 0.000 & 0.231 & 0.350 & 0.290 & 0.470 & 0.830 \\ 
  Gaussian & $-1.657 \cdot 10^{6}$ & 0.000 & 0.194 & 0.652 & 0.610 & 0.870 & 1.270 \\ 
   \bottomrule
\end{tabular}
}
\endgroup
\end{table}

\newpage
\begin{landscape}
\renewcommand{\arraystretch}{2}
\begin{table}[H]
\caption{Parameter values for the redistribution models, fit to data from study area \#2. The parameters $\rho$, $\rho_1$, and $\rho_2$ have units of kilometers; the remaining parameters are dimensionless.} 
\label{tab:redist_pars2}
\centering
\begingroup\fontsize{8pt}{9pt}\selectfont
\noindent\makebox[\textwidth]{%
\begin{tabular}{llll}
  \toprule
Kernel name & Tail type & $D(r) \propto$ & Max likelihood estimate \\ 
  \midrule
Pareto & Fat-tail & $\left(r + \rho\right)^{-(1+\nu)}$ & $\rho = 1.45 \cdot 10^{-2}, \;\nu = 1.61$ \\ 
  Student's $t$ & Fat-tail & $\left(1 + \frac{1}{\nu}\left(\frac{r}{\rho}\right)^2\right)^{-\frac{\nu + 1}{2}}$ & $\rho = 1.16 \cdot 10^{-2}, \;\nu = 1.47$ \\ 
  Bessel mixture & Mixture of Thin & $\theta \, K_{0}(\frac{r}{\rho_1}) + (1 - \theta)K_{0}(\frac{r}{\rho_2})$ & $\rho_1 = 2.65 \cdot 10^{-2}, \;\rho_2 = 7.84 \cdot 10^{-1}, \;\theta = 5.90 \cdot 10^{-1}$ \\ 
  Laplace mixture & Mixture of Thin & $\theta\exp\left(-\frac{r}{\rho_1}\right) + (1 - \theta)\exp\left(-\frac{r}{\rho_2}\right)$ & $\rho_1 = 1.55 \cdot 10^{-2}, \;\rho_2 = 4.91 \cdot 10^{-1}, \;\theta = 5.22 \cdot 10^{-1}$ \\ 
  Gaussian mixture & Mixture of Thin & $\theta\exp\left(-\left(\frac{r}{\rho_1}\right)^2\right) + (1 - \theta)\exp\left(-\left(\frac{r}{\rho_2}\right)^2\right)$ & $\rho_1 = 2.23 \cdot 10^{-2}, \;\rho_2 = 1.39, \;\theta = 4.40 \cdot 10^{-1}$ \\ 
  WMY & Thin tail & $\left(\frac{r}{\rho}\right)^{\kappa} K_{\kappa}(\frac{r}{\rho})$ & $\rho = 1.95 \cdot 10^{-1}, \;\kappa = 2.96 \cdot 10^{-8}$ \\ 
  Bessel & Thin tail & $K_{0}(\frac{r}{\rho})$ & $\rho = 2.15 \cdot 10^{-1}$ \\ 
  Laplace & Thin tail & $\exp\left(-\frac{r}{\rho}\right)$ & $\rho = 1.75 \cdot 10^{-1}$ \\ 
  Gaussian & Thin tail & $\exp\left(-\left(\frac{r}{\rho}\right)^2\right)$ & $\rho = 7.35 \cdot 10^{-1}$ \\ 
   \bottomrule
\end{tabular}
}
\endgroup
\end{table}
\end{landscape}
\renewcommand{\arraystretch}{1}

\section{Robustness check: inter-annual dispersal variability} \label{app:interannual}

\begin{figure}[H]
\centering
\makebox[\textwidth]{\includegraphics[scale = 1]{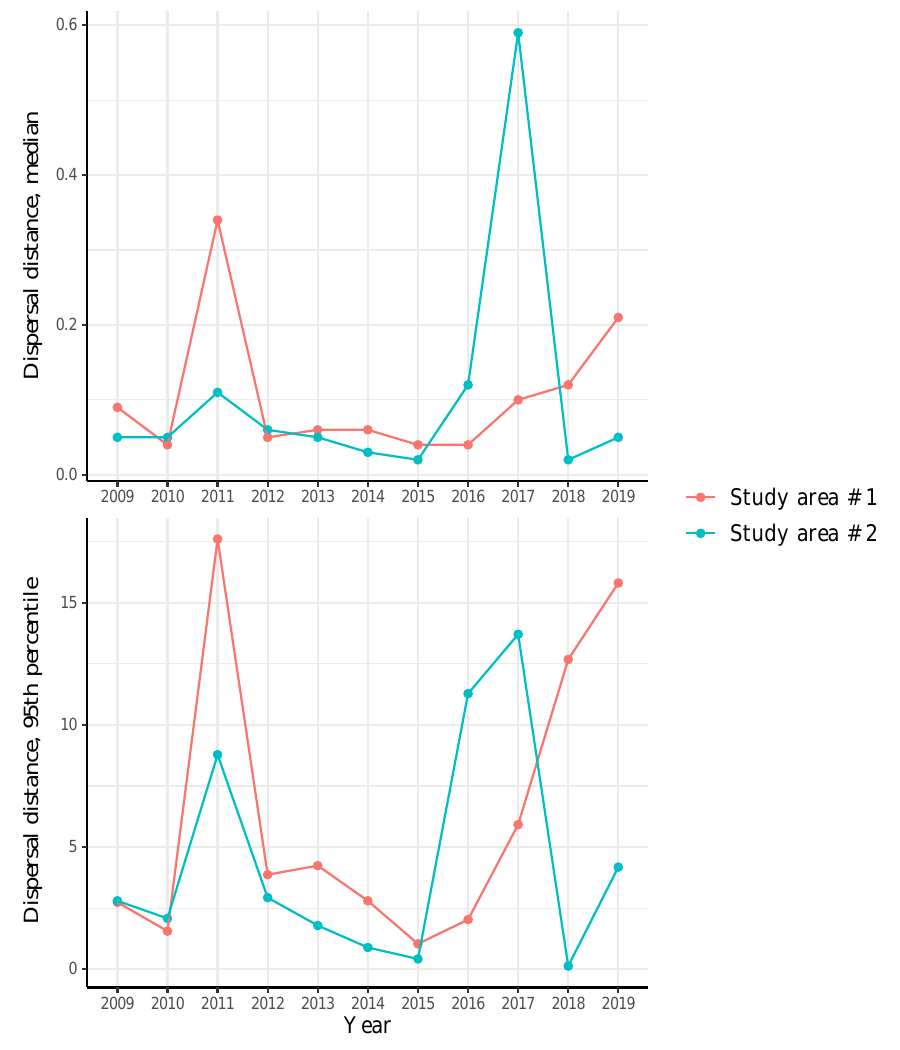}}
\caption{Estimates of dispersal distance across years, for both study areas.}
\label{fig:distance_time_series}
\end{figure}

Just as dispersal may exhibit spatial heterogeneity, it may also exhibit temporal heterogeneity. To examine this, we applied a Student's $t$ dispersal kernel to each year's data independently. The resulting estimates reveal that the median dispersal distance fluctuates over time, although generally within narrow bounds. Specifically, while the median typically stays in the range of 0.02 -- 0.1 km (centered around 0.05 km), there were 2/22 estimates which exceeded 0.3 km (Fig.~\ref{fig:distance_time_series}). 

\section{Additional information about characteristic length scales} \label{app:length}

Table \ref{tab.data} shows that the typical length scale of MPB dispersal varies widely across the literature, as well as data from related bark beetles. Here, in Table~\ref{tab.appdata}, we show in more detail where we obtained the stated values in that table for MPB specifically, and in Table~\ref{tab.appdata2} we show more detail for the stated values for related bark beetles.

\newpage 
\begin{landscape}
\renewcommand{\arraystretch}{2}
\begin{table}[H]
\caption{
Additional information on the stated values in Table~\ref{tab.data} for mountain pine beetle. This includes information about the study location as well as more details on where we obtain the characteristic scale of infestation.}
\label{tab.appdata}
    \centering
    \begingroup\fontsize{8pt}{9pt}\selectfont
    \noindent\makebox[\textwidth]{%
    \begin{tabular}{p{0.3\textwidth} l p{0.26\textwidth} p{0.6\textwidth}}
        \toprule \toprule
        Study & Scale (m) & Study location & Source of scale \\ \midrule \midrule
\citet{aukema_movement_2008} & 18 000 & British Columbia & 
Finds that the neighborhood of one cell is the most important predictor for infestations in 12 km x 12 km cells, which corresponds to roughly 18 km \\
\citet{koch_signature_2021} & 17 000 & British Columbia & Finds 17 km for the median dispersal distance from fitting an anisotropic WMY dispersal kernel to infestation data (\textit{personal communication}) \\
\citet{preisler_climate_2012} & 10 000 & Washington and Oregon & 
Finds that a distance weighted beetle pressure metric out to 10 km is one of the most important variables for all stages of outbreak \\
\citet{sambaraju_climate_2012} & 6000 & Western Canada &
Finds that the most important variable in a statistical model is infestation in the previous year within the 12 km x 12 km cells \\
\citet{howe_climate-induced_2021} & 5000 & British Columbia &
Finds that a distance weighted beetle pressure metric from the previous year within 5 km provided the best explanatory power for a statistical model fit to infestation data \\
\citet{carroll_assessing_2017} & 2000 & Alberta & 
Finds that more than 75\% of new infestations occur within 2 km of ``parent'' polygons containing the infestations from the previous year using heli-GPS data \\
\citet{simard_what_2012} & 2000 & Wyoming & 
Finds beetle pressure, up to 2 km, is the most important predictor for subsequent outbreaks for mountain pine beetle, spruce beetle, and Douglas-fir beetle \\
\citet{robertson_spatialtemporal_2009} & 1000 & Canadian Rocky Mountains &
Finds that most movements distances centre around 1 km using spatial–temporal analysis of moving polygons (STAMP) \\
\citet{strohm_pattern_2013} & 364 & Sawtooth National Recreation Area, Idaho & 
Finds MPB attacks should be spaced by 364 m by studying pattern formation in a model of chemical signalling including diffusion and chemotaxis and parameterized with data from \citet{biesinger_direct_2000} \\
\citet{powell_phenology_2014} & 5 -- 90 & Sawtooth National Recreation Area, Idaho & 
Finds that density dependent motility varies from an unimpeded motility of 3.79 km$^2$ per day to 18.5 m$^2$ per day in a fully stocked stand by fitting a phenology and dispersal model to aerial detection data \\
\citet{robertson_mountain_2007} & 30 -- 50 & British Columbia & 
Finds that the most common distances between newly and previously attacked trees are 30 m and 50 m given a search radius of 100 m from previously attacked trees \\
\citet{safranyik_dispersal_1992} & 30 & British Columbia & 
Finds that 86\% and 93\% of total captured beetles were found within 30 m of release site in a mark-recapture experiment in two years \\
\citet{heavilin_novel_2008} & 10 -- 15 & Sawtooth National Recreation Area, Idaho & 
Finds a mean and median dispersal distance of between 10 m to 15 m by fitting Gaussian and exponential dispersal kernels to aerial detection data \\ 
\citet{goodsman_aggregation_2016} & 10 & British Columbia and Alberta & 
Finds a median dispersal distance of 10 m by parameterizing a 2D dispersal kernel with mark-recapture data from \citet{safranyik_dispersal_1992} \\
\bottomrule
    \end{tabular}
}
\endgroup
\end{table}

\begin{table}[H]
\caption{
Additional information on the stated values in Table~\ref{tab.data} for related bark beetles. This includes information about the study location as well as more details on where we obtain the characteristic scale of infestation.}
\label{tab.appdata2}
    \centering
    \begingroup\fontsize{8pt}{9pt}\selectfont
    \noindent\makebox[\textwidth]{%
    \begin{tabular}{p{0.3\textwidth} l l l p{0.45\textwidth}}
        
        \toprule \toprule
        Study & Scale (m) & Study location & Bark beetle species & Source of scale \\ \midrule \midrule
        

\citet{withrow_spatial_2013} & 1000 -- 2500 & Colorado and Wyoming & Douglas-fir beetle & 
Finds average standard dispersal distances --- distances at which 68\% of infestations dispersed in a given year --- between 1000 m and 2500 m by quantifying the distance between infestations and modeling these distances using a cumulative Gaussian function\\
\citet{turchin_quantifying_1993} & 690 & Louisiana & 
Southern pine beetle &
Finds a median dispersal distance of 690 m for released beetles by fitting a dispersal model with mark-recapture data \\
\citet{werner_dispersal_1997} & 90 -- 300 & Alaska & Spruce beetle &
Finds that most recaptured beetles from standing trees dispersed between 90 m and 300 m in mark-recapture experiments\\
\citet{zumr_dispersal_1992} & 200 & Southern Bohemia & European spruce beetle &
Finds that most beetles (about 70\%) were captured within 200 m of release in mark-recapture experiments \\
\citet{dodds_sampling_2002} & 200 & Idaho & Douglas-fir beetle &
Finds that most beetles (over 90\%) were captured within 200 m of release in mark-recapture experiments  \\
\citet{kautz_quantifying_2011} & 100 & Bavaria & 
European spruce beetle &
Finds that 65\% of new infestations occurred within 100 m of the previous year's infestations \\ 
\citet{zolubas_recapture_1995} & 10 & Western Lithuania &
European spruce beetle & 
Finds that most beetles (about 67\%) were captured within 10 m of release in mark-recapture experiments for the second flight \\
    \bottomrule
    \end{tabular}
}
\endgroup
\end{table}

\end{landscape}

\section{Additional figures \& tables} \label{app:supp_fig}

\begin{figure}[H]
\centering
\makebox[\textwidth]{\includegraphics[scale = 1]{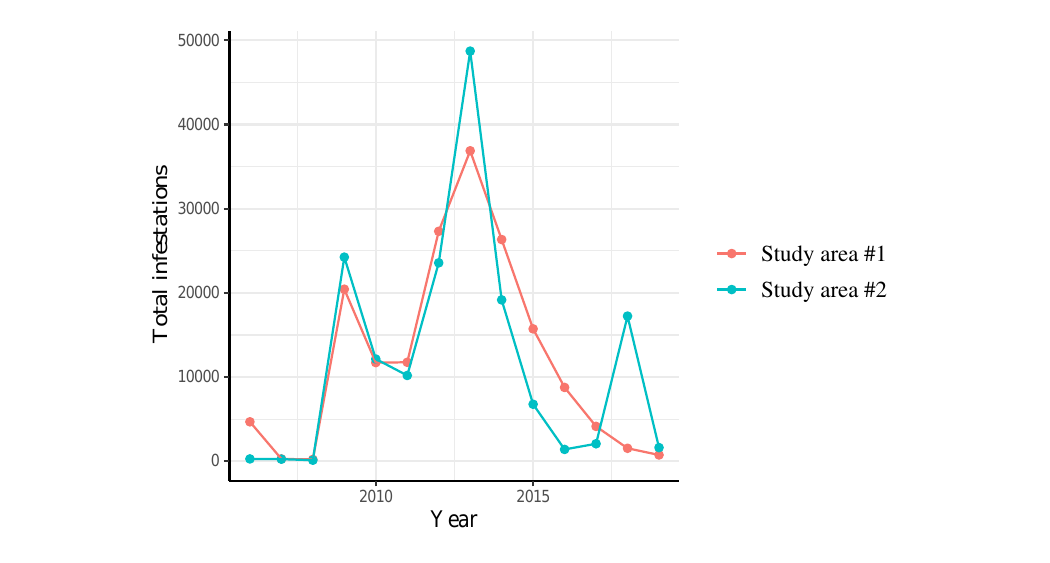}}
\caption{Time series of total infestations in both study areas. A large number of beetles were present from 2009--2019.}
\label{fig:infest_time_series}
\end{figure}

\begin{figure}[H]
\centering
\makebox[\textwidth]{\includegraphics[scale = 1]{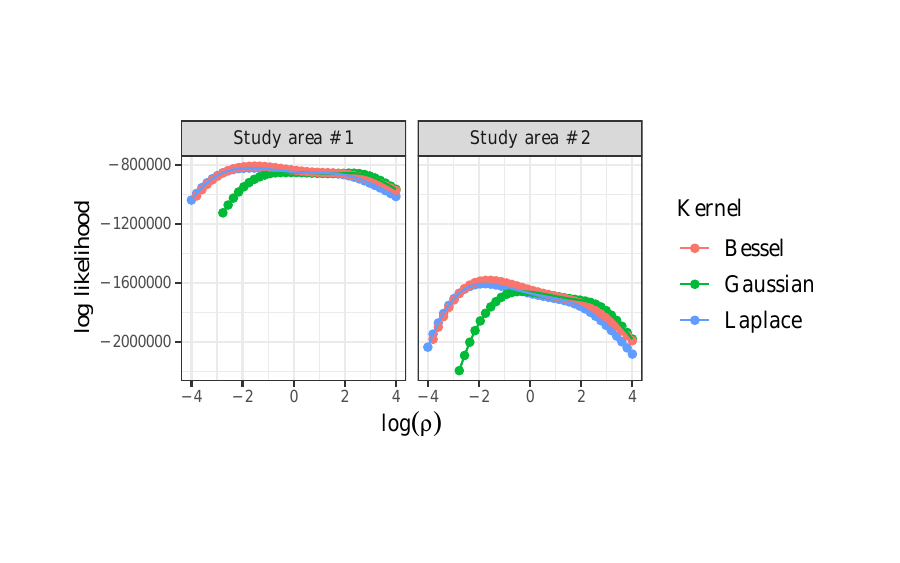}}
\caption{Likelihood profiles for the three one-parameter redistribution models. The gradient of the log likelihood is relatively flat for large values of the scale parameter $\rho$, which can cause the failure of gradient-based optimization methods. Instead, we used the grid search method to find the Maximum Likelihood Estimates for these models.}
\label{fig:ll_profiles}
\end{figure}

\begin{figure}[H]
\centering
\makebox[\textwidth]{\includegraphics[scale = 1]{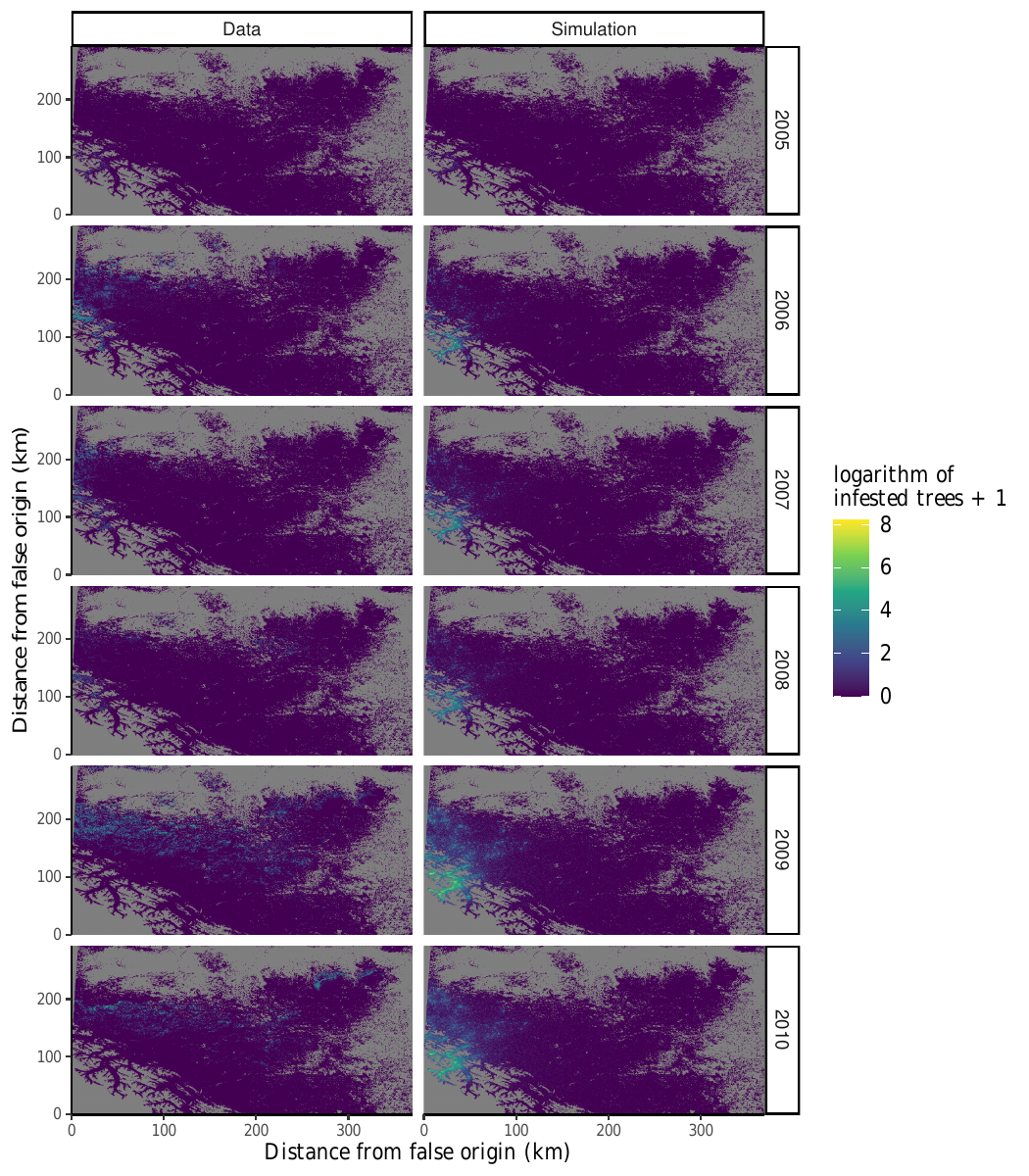}}
\caption{Comparison of the spatial distribution of infestations for both real and simulated data. Our simulations do not account for resource depletion, which leads to an unrealistic concentration of future infestations around the initial infestations. Our simulations do not account for beetle aggregation or clumpy dispersal, which leads to an unrealistically high spatial dispersion of infestations.}
\label{fig:sim_student_t}
\end{figure}

\end{appendices}

\bibliographystyle{apalike}
\bibliography{mpb_refs}

\end{document}